\documentclass[10pt]{elsart}
\usepackage{epsfig}
\usepackage{amsmath,amsfonts,amssymb}

\begin{document}

\begin{frontmatter}

\title{Detection of GRB signals with Fluorescence Detectors}

\author[lngs]{R. Aloisio},
\author[aq]{D. Boncioli},
\author[lngs]{A. F. Grillo},
\author[aq]{C. Macolino},
\author[aq]{S. Petrera},
\author[chi]{P. Privitera},
\author[aq]{ V. Rizi},
\author[aq]{F. Salamida}

\address[lngs]{INFN, Laboratori Nazionali del Gran Sasso, Assergi (L'Aquila), Italy}
\address[aq]  {Universit\'a de L'Aquila and Sezione INFN, L'Aquila, Italy}
\address[chi] {University of Chicago, Chicago, IL 60637, USA}

\begin{abstract}
Gamma Ray Bursts are being searched in many ground based experiments detecting the high energy component 
(GeV $\div$ TeV energy range) of the photon bursts. In this paper, Fluorescence Detectors are considered as 
possible candidate devices for these searches. It is shown that the GRB photons induce fluorescence emission of UV photons on a wide range of their spectrum. The induced fluorescence flux is dominated by GRB photons from 0.1 to 
about 100 MeV and, once the extinction through the atmosphere is taken into account, it is distributed
over a wide angular region. This flux can be detected through a monitor of the diffuse photon flux, provided that its maximum value exceeds a threshold value, that is primarily determined by the sky brightness above the detector. The feasibility
of this search and the expected rates are discussed on the basis of the current GRB observations and the existing
fluorescence detectors.
\end{abstract}

\begin{keyword}

Gamma Ray Bursts, GRB, Fluorescence.

\end{keyword}

\end{frontmatter}

\section{Introduction \label{Intro}}

Almost forty years after their discovery \cite{GRB} Gamma Ray Burst (GRB) remain one of the most interesting objects in High Energy Astrophysics. GRB are characterized by an intense emission of gamma rays with a very short time duration (in the range of 0.1 up to 100 seconds). Even though the emission mechanism of GRB is still not well understood, theoretical models aiming to explain this emission share common features with gamma-rays produced by synchrotron and inverse Compton (IC) emission of charged particles accelerated in the shock wave of a fireball, the relativistic shock produced by an unknown catastrophic event: most probably coalescence of compact objects (short bursts) and gravitational supernovae such as type Ib and II (long bursts).

The successful observations of the BATSE instrument, onboard the Compton Gamma Ray Observatory (1991-2000) 
\cite{BATSE}, and the BEPPO-SAX satellite (1997-2002) \cite{BEPPO} have opened the possibility of GRB phenomenology with the collection of a large data set. These observations performed at different frequency bands prove that the GRB sources are at cosmological distances with a typical isotropic emission in the range of $\rm 10^{51}\div 10^{54}$ ergs, that makes GRB the most powerful sources in the universe. Nowadays the detection of GRBs is performed by HETE, Integral and Swift satellites \cite{HETE,INTEGRAL,SWIFT}.

The observations performed so far are concentrated in the keV-MeV energy range, where the observed emission has a spectrum in agreement with synchrotron based models. An important step forward in unveiling the GRB emission mechanisms and its hidden engine is the extension of the observations to the highest energies in the GeV-TeV range. High energy emission from GRB has been detected in several GRBs \cite{Gonzales2003,Hurley1994,Atkins2003}. In the framework of the fireball model TeV photons can be produced in both internal and external shock of the GRB 
\cite{Dai2002,Razzaque2004,Zhang2001,Bottcher1998} and either through electrons IC scattering or protons synchrotron emission. In \cite{Wang2001} it is proposed that the cross IC between photons and electrons in forward and reverse shocks produces TeV photons. TeV photons can also be produced by IC scattering of the shock accelerated electrons in the external shock off a bath of photons that overlaps the shocked region. This photon bath can be either the prompt gamma-ray emission or the late time X-ray emission. On general grounds, if the high energy emission of GRB is dominated by IC scattering the typical energy separation between synchrotron and IC emissions is  of the order of $\gamma_e^2$, being $\gamma_e$ the electrons Lorentz factor which is typically $\rm 10^2\div 10^3$ \cite{Zhang2001}, in this way a GRB can easily emit photons exceeding the TeV energies.

The observation of GRBs at either low and high energy up to 300 GeV will be performed by the GLAST 
satellite, the next generation GRB dedicate satellite that will be launched in the beginning of 2008. Nevertheless, up to now the only way of observing the high energy emission from GRBs is through ground based experiments that are less constrained in size and, in principle, can provide observation up to extreme energies in the TeV range. Ground detectors measure the secondary particles produced by the interaction of $\gamma$-rays with the atmosphere (photons, $\rm e^{\pm}$, Cerenkov and fluorescence light) that reach the ground.

Among different ground based techniques the Cerenkov light detection has a small field of view (only about few square degrees), this makes Cerenkov telescopes less suitable for the observation of transient and unforeseeable events such as GRB. Nevertheless, triggering the Cerenkov telescope with alerts by satellites it is possible to observe late emissions from GRB. Recently the MAGIC telescope allowed the observation of nine different GRBs as possible sources of high energy $\gamma$-rays, these observations were performed using the alerts from Swift, HETE-II, and Integral \cite{Magic}.

Air shower arrays and fluorescence telescopes because of their large field of view (almost $\pi$ sr) are better suited to observe high energy emission from GRBs. In this respect some indications of TeV emission from GRBs at low red-shift were provided by several ground based experiments as MILAGRO \cite{Milagro}, HEGRA \cite{Hegra} and TIBET \cite{Tibet}. These observations are based on the "single particle technique" \cite{Vernetto},  that consists in an increase of the background signal in all on ground detectors on a time scale corresponding to the burst duration. Recently the Auger collaboration proposed an analysis campaign to test the single particle technique on the surface detector of the Auger observatory \cite{Xavier}.

In the present paper we will concentrate our attention on the fluorescence emission produced by high energy photons from GRBs. Photons with energy larger than $\rm 0.1\div 1.0$ MeV, interacting with the atmosphere, produce an isotropic fluorescence emission through Compton scattering as well as (at $\rm E>E_c=80$ MeV) electromagnetic cascades. As we will discuss, the fluorescence emission due to the Compton scattering represents a possible advantage of the fluorescence detection, because it opens the GRB ground based observations to the energies of the peak GRB emission (in the $\rm 100$ keV range). 
In the following we will determine the expected emission discussing
the detection capabilities of fluorescence detectors. 

The paper is organized as follows. In section \ref{UV} and \ref{MC} we
will study the fluorescence emission in the atmosphere using an
analytic technique as well as a Monte Carlo approach. In section
\ref{FDground} we discuss the sensitivity of fluorescence instruments
at ground. Finally
 conclusions take place in section \ref{Con}. 

\begin{figure}[t!]
\centering
\includegraphics*[width=.6\textwidth]{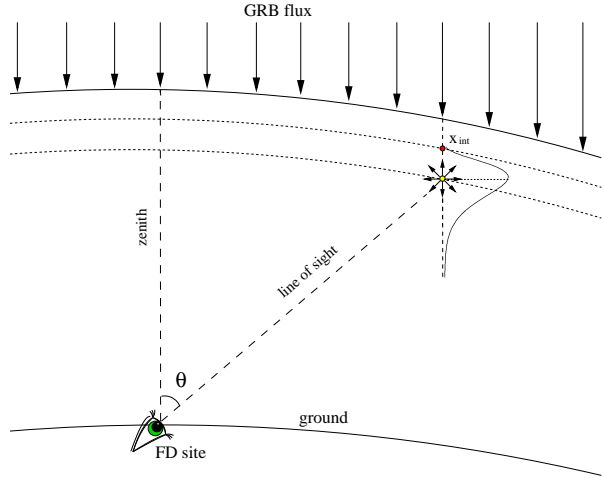}
\caption {\label{fig.scheme}{\small{Scheme of a GRB arriving on the 
      Earth. The GRB flux can be seen as a parallel photon beam hitting the top of the
      atmosphere. The figure shows a photon, with energy higher than E$_c = 80$ MeV,
      interacting with air molecules and producing an electromagnetic cascade.}}}
\end{figure}
 
\section{The fluorescence emission from GRB in the atmosphere
  \label{UV}}

Individual photons hitting the atmosphere with energies in the range
from MeV to TeV are absorbed through two concurring processes: Compton
scattering and pair production (with a subsequent cascade
development). In both processes the excitation of nitrogen molecules
by electrons and positrons gives rise to a fluorescence emission in
the visible-UV frequency band. 
This emission does not produce signals observable in fluorescence
detectors, whose  trigger threshold 
for hadron showers is of the order of 10$^{17}$ eV. On the other hand these 
detectors monitor routinely the background flux \cite{AugerMonitor} to
control  the detector response and prevent possible damages to the
apparatus. 
This flux is sampled with time periods usually long (few to tens
  of seconds) with respect to the development of the showers, but
  roughly of the same size as the duration of the GRB. Therefore we
  can expect that the GRB induced fluorescence emission produces an
  increase of the measured background. The aim of the present paper 
is to estimate if such an increase is detectable under reasonable assumptions on the GRB fluxes.

In the following we will assume that the GRB source is placed at the zenith of the fluorescence detector.  As described in figure \ref{fig.scheme}, the photon burst is viewed by an on ground observer as a beam parallel to the zenith axis spread over the whole atmosphere above the detector. Primary photons penetrate the atmosphere up to a depth determined by their absorption coefficient: Compton scattering and pair production are the two concurring processes contributing to photon interaction in the atmosphere. 

\begin{figure}[t!]
\centering
\includegraphics*[width=.75\textwidth]{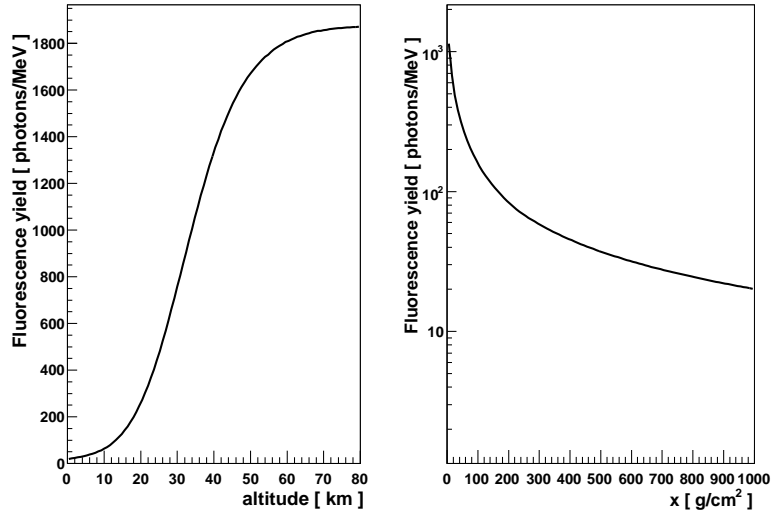}
\caption {\label{fig.FYAF}{\small{Fluorescence yield per unit energy
      deposit as a function of the altitude (left) and grammage (right), from a parameterization described in the text.}}}
\end{figure}

The first step is the calculation of the fluorescence emissivity, the rate of fluorescence photons emitted per unit volume and unit solid angle, at the air layer having grammage x. This is determined by photons interacting at x$_{int}$, above the emission point x (see  figure \ref{fig.scheme}). For most of the photon energies the path-length of charged secondaries, which induce fluorescence emission, is small compared with the altitude and it can be reasonably assumed that x = x$_{int}$. For energies exceeding the critical energy (E$_c = 80$ MeV) photons develop into electromagnetic showers, whose longitudinal size increases logarithmically with the energy. For this reason we will treat separately the two processes distinguishing between the two energy regimes of primary photons. Common assumptions in the calculation are:

\begin{itemize}
\item{Curved Earth
\begin{equation}
h \simeq r~ cos \theta~+~\frac{r^2~sin^2 \theta}{2~R_{\oplus}}
\label{rvsh}
\end{equation}
with r and h being the distance and height from the fluorescence
emission source, $\rm \theta$ the observation zenith angle and $\rm R_{\oplus}$ the Earth radius.}
\item Grammage vs height dependence as for the U.S. standard atmosphere \cite{USSt}.
\item The absolute fluorescence yield of the 337~nm band in air was taken to be 5.05 photons/MeV of energy deposit at 293~K and 1013~hPa, derived from~\cite{naganoFY}. The wavelength and pressure dependence of the fluorescence spectrum was obtained from the AIRFLY experiment \cite{AirFly}. The corresponding fluorescence yield is shown in figure \ref{fig.FYAF}.
\item The high energy GRB flux on Earth is assumed as 
\begin{equation}
F(E)=F_0 \left (\frac{E}{{\rm MeV}} \right)^{-\alpha}
\label{Eflux}
\end{equation}
being $\rm F_0$ a normalization factor depending on the relevant GRB parameters such as its luminosity and redshift. The flux is assumed to extend from $\rm E_{pk}$, the so-called synchrotron peak, to the maximum injected energy 
$\rm E_{max}$. We set $\rm E_{pk}$ = 100 keV, and $\rm E_{max}$ ranging from 100 GeV up to 100 TeV.
\end{itemize}

\subsection{Low energy (E $<$ E$_{c}$)
  \label{Compt}}

At low energy $\rm E<E_c=80$ MeV, fluorescence emission is mainly
produced by the Compton scattered electrons.  From energies above 10
  MeV pair production contributes as well and finally dominates as energy
  approaches the critical energy. For both processes the energy of primary and secondary
  electrons is released at short distance from the photon interaction
  point. In our calculation we simply assume that the emission is
  concentrated at the photon interaction point. 
In this case, the emissivity at grammage x can be simply obtained convoluting the GRB flux with the photon interaction probability and the fluorescence conversion factor: 
$$I(x)~=~\frac{\rho}{4 \pi}~ \int_{E_{pk}}^{E_{c}}
dE ~F(E) ~ P(x_{int},E) ~E~Y$$
\begin{equation}
=~\frac{\rho(x)~Y(x)}{4 \pi}~ \int_{E_{pk}}^{E_{c}}
dE ~E~F(E) ~ \mu(E)~exp[-\mu(E)~ x]
\label{emissC}
\end{equation}
where $\rho$ is the air density, Y is the fluorescence photon yield in photons per energy deposit and $\mu$ is the total photon absorption coefficient.

\subsection{High energy (E $>$ E$_{c}$)
\label{EMshow}}

Photons above critical energy produce e.m. showers. For simplicity we assume that even in this case photon emission is point-like, but at $\rm x_{int}+X_{max}$, being $\rm X_{max}(E) = X_0~log(E/E_{c})/log2$ the grammage at the shower maximum and $\rm X_0 = 37 ~g/cm^2$ the air radiation length. In the next section we will release this assumption in a comparative Monte Carlo simulation. Therefore we get:
$$I(x)~=~\frac{\rho}{4 \pi}~ \int_{E_{c}}^{E_{2}(x)}
dE ~F(E) ~ P(x_{int},E) ~E~Y$$
\begin{equation}
=~\frac{\rho(x)~Y(x)}{4 \pi}~ \int_{E_{c}}^{E_{2}(x)}
dE ~E~F(E) ~ \mu(E)~exp[-\mu(E) (x - X_{max}(E))]
\label{emiss}
\end{equation}
where $\rm E_{2}(x) = min[E_{max} ,~ E_{c} ~exp(\frac{log 2 ~ x}{X_0})].$

\begin{figure}[t!]
\centering
\includegraphics*[width=.75\textwidth]{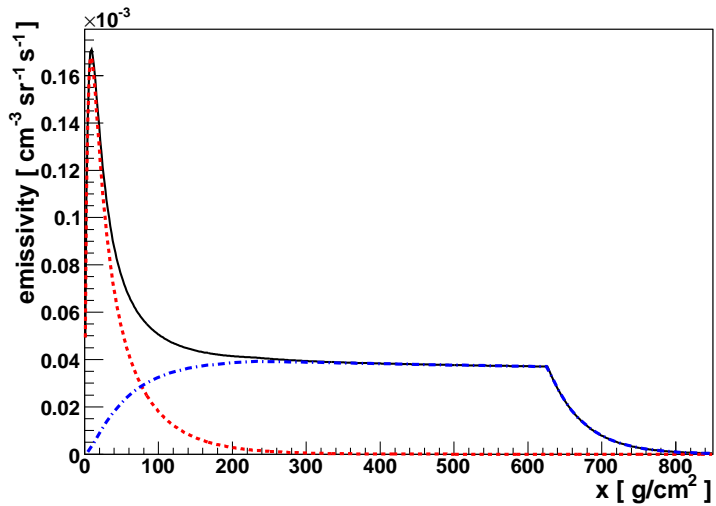}
\caption {\label{fig.emiss}{\small{Fluorescence emissivity as a
      function of the atmosphere grammage for a GRB at
      the zenith with a flux having $\rm F_0~ =~ 1~MeV^{-1} cm^{-2} s^{-1}$
      and a spectral index 2. Solid, dashed and dot-dashed lines refer to total, low and
      high energy respectively. Low energy emissivity is calculated from equation 
      (\ref{emissC}), with E$_{pk}$ = 100 keV. High energy emissivity is calculated from 
      equation (\ref{emiss2}), with E$_{max}$ = 10 TeV.}}}
\end{figure} 

It has to be noticed that the photon absorption coefficient varies very mildly in the energy range of our calculation: it is about 20 $\%$ below its asymptotic value at $\rm E_{c}$ reaching it at few GeV. Therefore it can be reasonably assumed constant and equal to $\rm 7/(9 X_0$) so that we can integrate analytically eq. (\ref{emiss})
\begin{equation}
I(x)=~\frac{F_0}{4 \pi}~\frac{7}{9 X_0}~ \rho(x)~Y(x)~E^{2}_{c} \left(\frac{MeV}{E_{c}}
\right)^\alpha~exp \left( - \frac{7 x}{9 X_0}\right)~f(x)
\label{emiss2}
\end{equation}
with:
$$f(x) = \frac{1}{2-\alpha+\frac{7}{9 log 2}} \times
\left\{\begin{array}{c}
\left( exp\left[log 2 ~(2-\alpha+\frac{7}{9 log
        2}) \frac{x}{X_0} \right]~-~1 \right) \qquad  x < X_{max}(E_{max}) \\
\left( \left[ \frac{E_{max}}{E_{c}} \right]^{(2-\alpha+\frac{7}{9 log 2})}~-~1 \right) ~~~~~~~~~~~~~\qquad x >  X_{max}(E_{max})
\end{array}\right.$$

Figure \ref{fig.emiss} shows the total, low and high energy fluorescence emissivities as a function of the atmosphere grammage for a GRB flux with $\rm F_0~ =~ 1~MeV^{-1} cm^{-2} s^{-1}$ and a spectral index 2~\footnote{The case of $\rm \alpha= 2$ is of particular interest because the high energy emissivity shows a wide flat-top from x $\simeq$ 200 g/cm$^2$ up to  $\rm X_{max}(E_{max})$. In fact for such spectral index I(x) is simply proportional to $\rm \rho(x) \cdot Y(x)$ for $\rm x < X_{max}(E_{max})$, because of a cancellation of the two exponentials in eq. (\ref{emiss2}).}. The high energy curve  refers to eq. (\ref{emiss2}). Once using eq. (\ref{emiss}) and the actual $\rm \mu(E)$ dependence the curves differ by few percent at low x.

\begin{figure}[t!]
\centering
\includegraphics*[width=.70\textwidth]{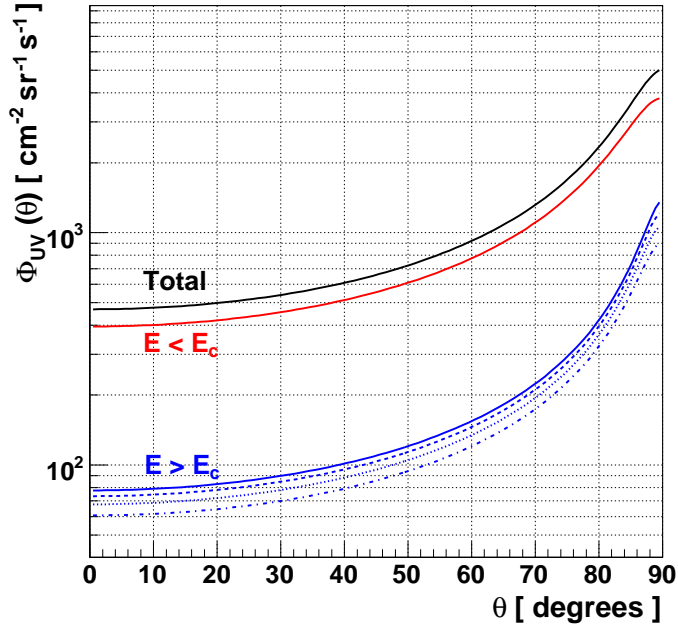}
\caption {\label{fig.flux}{\small{Emitted fluorescence photon flux as a function of the line of sight zenith angle. The red line refers to low energy. The solid, dashed, dotted and dot-dashed blue lines refer to the high energy flux with $\rm E_{max} =$ 100, 10, 1 TeV and 100 GeV respectively. The black line is the total fluorescence flux for $\rm E_{max} =$ 10 TeV.
The flux assumptions  are the same as in figure \ref{fig.emiss}.}}}
\end{figure}

\subsection{The fluorescence photon flux
  \label{UVflux}}

Once determined the emissivity the expected flux on a fixed line of sight is, by definition, the integral over the line of sight of the emissivity itself. Fixing a generic line of sight at angle $\rm \theta$ the fluorescence flux will be
\begin{equation}
\Phi_{UV} (\theta) = \int_{0}^\infty I(r)~dr
~ =~ \int_{0}^{x_{G}} I(x)~\frac{\partial r}{\partial h}(h(x),\theta)~ \frac{dh}{dx}~{dx}
\label{PhiUV}
\end{equation}
where $\rm r(h,\theta)$ is obtained inverting eq. (\ref{rvsh}) and dh/dx is obtained from the U.S. standard atmosphere model.

In figure \ref{fig.flux} we show the fluorescence flux due to the low and high energy regimes 
and taking different values for the maximum allowed energy of photons $\rm E_{max}$.
It can be recognized that increasing the zenith angle of the line of sight the fluorescence flux increases and its value is weakly dependent on the maximum energy of primary photons. The increase in the flux with the line of sight zenith angle can be understood taking into account that with the zenith angle increases the portion of the emitting atmosphere that contributes to the flux. 

\begin{figure}[t!]
\centering
\includegraphics*[width=.65\textwidth]{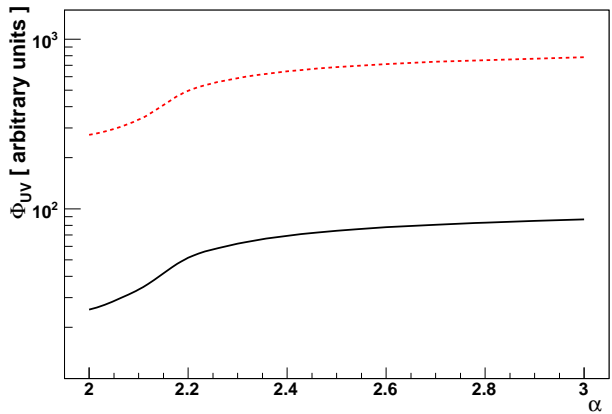}
\caption {\label{fig.features}{\small{Dependence of the vertical (solid line) and
horizontal (dashed line) photon flux on the spectral index.}}}
\end{figure}

In figure \ref{fig.features} we describe the dependence of the fluorescence flux on the spectral index of the primary photons. The two curves refer to the two cases of a vertical line of sight ($\theta=0$) and an horizontal one ($\rm \theta=\pi/2$). It is interesting to note how the flux slightly increases with the spectral index, this is a direct consequence of the dominance of the fluorescence production through Compton scattering (see figure \ref{fig.flux}) which is effective at low energy. Increasing the power law index the number of low energy photons is increased and the Compton scattering emission becomes more effective.  

\begin{figure}[t!]
\centering
\includegraphics*[width=.75\textwidth]{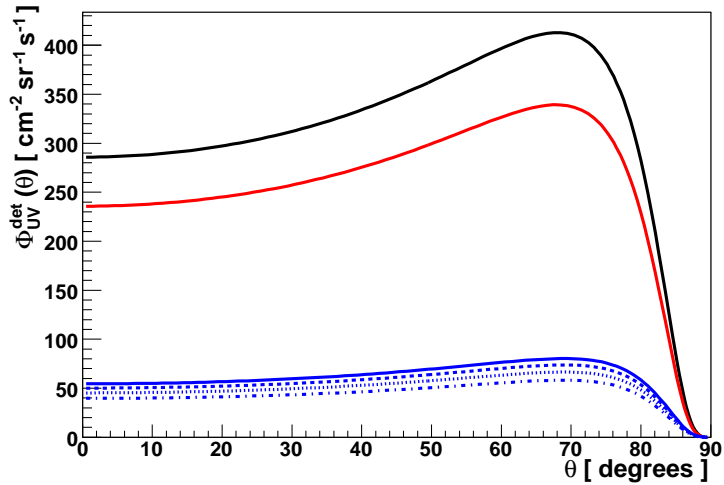}
\caption {\label{fig.detflux}{\small{Fluorescence photon flux at ground as a
      function of the zenith angle. The line features are the same as in fig. 
\ref{fig.flux}. The flux assumptions  are the same as in figure \ref{fig.emiss}.}}}
\label{fig.fluxUV}
\end{figure}

 Finally we have to address the issue of the extinction effect due to the atmosphere in the propagation of fluorescence photons. This effect is important and could sensibly reduce the observed flux in particular in the case of lines of sight with high zenith angle. 
To account for the extinction effect we adopted a simple model by Garstang \cite{Garstang}, which reproduces the night-sky brightness at several observatories and sites. In figure \ref{fig.detflux} the expected flux at ground is shown for the same assumptions of figure \ref{fig.flux}. It can be seen that the extinction is increasingly effective for lines of sight with high zenith angle, reducing considerably the angular dependence shown in figure \ref{fig.flux}.

\section{Monte Carlo simulation \label{MC}}

In this section we will determine the fluorescence emissivity and the corresponding flux with
the help of a Monte Carlo (MC) simulation. In this way we will test the assumptions we made
in the analytical calculation (AC) of the fluorescence emissivity and in particular the
hypothesis of point-like fluorescence emission.
The MC simulation sample consists of millions of photons hitting the top of the atmosphere.
These photons are uniformly distributed over an area of $\rm 1.2\times10^{6}$ km$^{2}$. 
This area has been chosen intersecting the horizon plane with the spherical surface 
corresponding to an altitude at which the residual atmospheric density
is about 10 $\rm g/cm^2$. 
Photon energies are generated according to the GRB photon flux, 
we restrict here our analysis to the case
of $\alpha=2$.

Each photon is individually followed through the atmosphere and the
interaction point $\rm x_{int}$ is 
generated according to $\rm \mu(E) ~exp[-\mu(E) x]$, where the absorption coefficient $\rm \mu(E)$
takes into account all possible production processes (i.e Compton, pair, photoelectric, $\rm \mu$ pair
and photo-nuclear).
In order to check the AC result of Sect. \ref{UV}  we have first used the same assumptions with the
fluorescence emission concentrated at a single point, $\rm x_{int}$
for $\rm E < E_{c}$ and $\rm x_{int} +  X_{max}$ for $\rm E > E_{c}$. In the
latter case we used
$\rm X_{max} = X_0 log(E/E_{c})/log2$ according to the Heitler model \cite{Heitler} as already done in Sect. \ref{UV}.

\begin{figure}[t]
\centering
\includegraphics*[width=.75\textwidth]{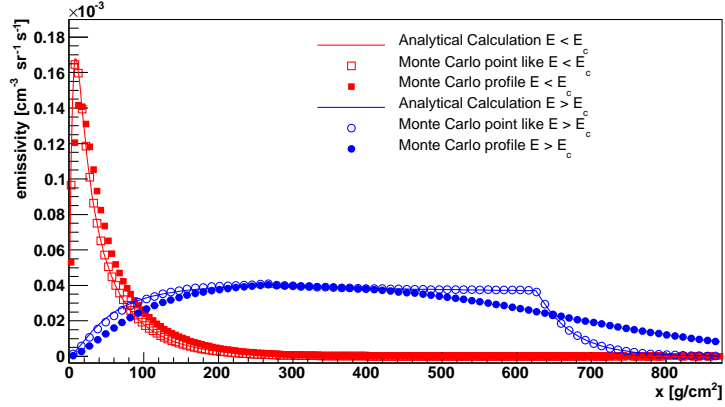}
\caption {\label{fig.compMC}{\small{Comparison between the
      emissivity calculated with the analytical model (solid lines) and
      Monte Carlo simulation using both the point like emission at the
      shower maximum (open markers) and the distributed
      emission (solid markers).}}}
\end{figure}

The result of our simulation is shown in figure \ref{fig.compMC}. 
The open squares and circles represent the MC simulation, 
respectively for $\rm E < E_{c}$ and $\rm E > E_{c}$ while the solid lines
refer to the AC. 
The two results are in perfect agreement as expected.

Releasing the assumption of a point-like emission, we need to take into account the longitudinal
development of secondaries. In the energy range $\rm E < E_{c}$ Geant4 \cite{Geant4} was used to simulate a
library of energy deposit profiles, while for $\rm E > E_{c}$ the shower development was parameterized
according to \cite{Profile}. For each point of the shower profile the energy deposit is evaluated and
converted into fluorescence photons (see figure \ref{fig.FYAF}). In this way each event contributes to the
emissivity in a finite interval of grammages. The result, represented by the solid squares and circles, 
is shown in figure \ref{fig.compMC}.
In the energy range $\rm E < E_{c}$ the effect is a smoothing of the peak
resulting from the AC and a slight shift to deeper grammages.
In the region $\rm E > E_{c}$ the smoothing effect is greater with respect to the lower energy range. 

\begin{figure}[t]
\centering
\includegraphics*[width=.75\textwidth]{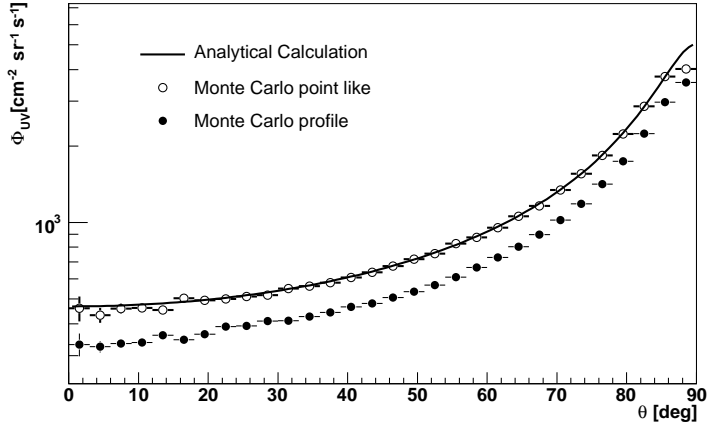}
\caption {\label{fig.fluxMC}{\small{Comparison between the
      GRB fluorescence flux calculated with the analytical model (black line) and
      Monte Carlo simulation using both the point like emission at the
      shower maximum (black open circles) and the distributed
      emission (black solid circles).}}}
\end{figure}

In spite of the differences in the emissivity curves the resulting emitted
fluorescence fluxes, for the MC
and AC, are similar. The relative difference between the MC
and AC flux depends on the zenith angle. It is about 30$\%$ at the
zenith
and decreases with the angle.  
In figure \ref{fig.fluxMC} the two emitted fluxes for the MC and AC are
shown. For this comparison 
the effect of the atmospheric transmission is not considered.

\section{Sensitivity at ground \label{FDground}}

In this section we will address the GRB detection capabilities of a fluorescence detector at ground. Hereafter we will restrict our considerations to the case of a spectral index $\alpha=2$, synchrotron peak energy $\rm E_{pk}= 0.1$ MeV and $\rm E_{max}=10$ TeV. The same results for fluxes with other values of the spectral index $\alpha$ can be easily derived from figure \ref{fig.features}. 

In figure \ref{fig.detflux} one can see that the fluorescence flux at
ground is at most 400 photons $\rm cm^{-2} sr^{-1} s^{-1}$. This value
corresponds to a GRB flux constant $\rm F_0$ = 1 $\rm MeV^{-1} cm^{-2} s^{-1}$. 

The detectability of the induced fluorescence flux is primarily limited by the level of the UV background flux at the
fluorescence detector site. This flux has been measured at the Auger site \cite{Augerbkgr} and is of the order of:
$$\Phi_{bkgr}~\simeq~\frac{100}{m^2 ~deg^2 ~\mu s}~\simeq~
\frac{3.3~10^7}{cm^2~ sr ~s}$$
with a mild dependence on the zenith angle. It is sensible to infer that it is roughly the same in all the sites suitable for fluorescence detection.

Assuming to be able to detect a GRB signal when an extra flux of $\rm \delta \times \Phi_{bkgr}$ is added up to the background, a signal is detectable if the GRB flux constant $\rm F_0$ is:
\begin{equation}
F_{0}~\gtrsim~\delta \times \left( \frac{\Phi_{bkgr}}
{400~cm^{-2} sr^{-1} s^{-1}} 
\right)~MeV^{-1} cm^{-2} s^{-1}~\simeq~
\delta \times 8~10^4~MeV^{-1} cm^{-2} s^{-1}
 \label{F0min}
\end{equation}
The lower limit given in eq. (\ref{F0min}) applies if the burst duration $\rm \Delta t_b$ exceeds the time window $\rm \Delta t$ used to measure the background flux. For shorter bursts the GRB signal is diluted into this window and the threshold 
(\ref{F0min}) is raised by a factor of about $\rm \Delta t /\Delta t_b$. The limit on the flux constant is easily converted into a limit on intensity, the fluence $\mathcal{F}$ in the time window $\rm \Delta t$. In fact:
\begin{equation}
\mathcal{I}(E_1,E_2)~=~\frac{\mathcal{F}}{\Delta t} ~\gtrsim~ 0.13 \times \frac{\delta}{\epsilon}
\left(\frac{erg}{cm^2~s}\right)~
\int_{E_1}^{E_2} \frac{d E}{E}
 \label{Fluence}
\end{equation}
where $\rm \epsilon = min [1,\Delta t_b /\Delta t ]$.
In order to compare such intensities with observations, one has to extend the integral limits to the range of the observations. We used Swift \cite{SWIFT} as a reference where fluences are measured in the interval 15$\div$150 keV. Then we constrained eq. (\ref{Fluence}) from our minimum energy, 100 keV, up to 150 keV. Denoting such intensity by $\mathcal{I}_{100}$ one gets the detectability condition:
\begin{equation}
\mathcal{I}_{100} ~\gtrsim~ 0.05\times \frac{\delta}{\epsilon}~\left(\frac{erg}{cm^2~s}\right)
 \label{Ilim}
\end{equation}

From the Swift GRB Table \cite{SWTable} we retrieved the fluences $\mathcal{F}$ and T90 durations of 272 GRBs. The corresponding average intensities $\mathcal{I}_{100}$ have been deduced by a simple model of the
signal\footnote{The GRB flux was assumed to be represented by a broken power law having a break at $\rm E_{pk} = 100$ keV and spectral indexes equal to -1 and -2 before and after the break. Then one gets $\mathcal{F}_{100} = 0.82 ~\mathcal{F}$, $\mathcal{F}$ being the fluence retrieved from the Swift catalogue.} and then fitted by a simple power law:
\begin{equation}
\frac{dN}{d\mathcal{I}_{100}}~\propto~ 
 \mathcal{I}_{100}^{~-\beta}
\label{IFunc}
\end{equation}
getting $\rm \beta = 1.59 \pm 0.07$.

\begin{figure}[t!]
\centering
\includegraphics*[width=.75\textwidth]{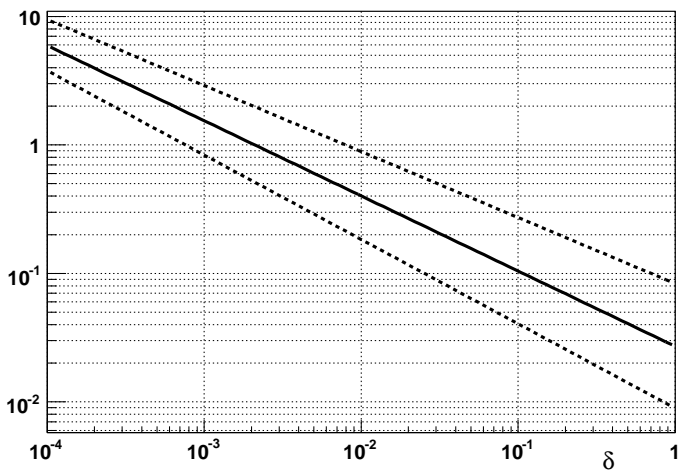}
\caption {\label{fig.GRBRate}{\small{Number of expected GRBs in 10
      years from fluorescence detection as a function of the threshold
      parameter $\delta$. The solid line corresponds to the best fit
      of $\beta$ with its one sigma deviation (dashed line), the case $\epsilon=1$ 
      is assumed.}}} 
\end{figure}

Using eq. (\ref{Ilim}) and assuming for Swift an effective exposure of
2$\pi\times$ 30 $\% \times$ 3 yr and a duty cycle of the fluorescence
detector of 10\% one gets the expected event rate as a function of the
threshold parameter $\delta$. This is shown in figure
\ref{fig.GRBRate}. 
It can be recognized that to achieve the
sensitivity of detecting at least one event in 10 years one needs a
threshold $\delta$ not exceeding one percent under the most favourable
conditions of current GRB estimates. More realistically one needs
$\delta$ to be about 1 per mill or better.

A discussion about the ability of reaching such challenging threshold is beyond the aim of the present paper.
From eq. (\ref{Ilim}) we can simply point out what are the minimal
requirements for a fluorescence detector aiming to detect GRB
signals. These are: 
\begin{enumerate}
\item capability of detecting very low intensity excesses with respect
  to the background, in order to get $\delta$ of the order of one per mill
  or even better;
\item wide (possibly full) sky coverage through detector pixels;
\item short sampling period of the photon intensity, in order to get 
the GRB signal spread out
  over several time bins;
\item full-time measurement of the light intensity, in order to
  maximize the number of photons at each sample.
\end{enumerate}
It can be noticed that the first three requirements are correlated one with each other.  In fact
the possibility of detecting a true signal and the minimum threshold applicable to data rely on the collective effects that
a GRB induces in the FD detector, i.e. the number of pixels jumping at
the same time (item 2) and the number of contiguous time bins realizing a minimum
pixel majority (item 3). In fact, the higher are such multiplicities
the lower is the probability for the detected signal of being
generated by random noise. Consequently these conditions allow to set very
low thresholds.

Looking at current detectors one can see that only a part of these
requirements are fulfilled. In the case of the FD system of the Pierre
Auger Observatory \cite{PAO}, requirement (2) is fulfilled
because of a very fine pixellation (about 10,000 pixels with
1.5$^\circ$ field of view each) and a coverage of about half of
the whole sky. But the other parameters are not adequate for a GRB
search. In particular the sky brightness is sampled with a 30 s period,
which is too long for signal sequencing. Moreover the light
intensity is measured (through ADC variances) in a 6.5 ms time
window for each sample, being the monitor system required to
detect an intensity increase of 5 $\%$ or more \cite{stat}. This window allows the collection of a very small fraction (about 2 per mill) of the total signal.

\section{Conclusions \label{Con}}

The fluorescence emission induced in the atmosphere by photons from Gamma Ray Bursts can be
detected by the fluorescence detectors used in UHECR physics. This detection consists in an increase of the diffuse sky brightness in a way not different than the single particle technique used to detect GRB signals in surface detectors \cite{Vernetto,Xavier}.

The UV photon emissivity in the atmosphere and the corresponding flux produced by a GRB have been calculated using an analytic technique as well as a Monte Carlo approach. It is important to notice that the GRB photons induce fluorescence emission on a wide range of their spectrum, with the higher contribution provided by lower energies
from 0.1 to about 100 MeV. This emission represents an important
difference with the surface detection of GRB that is not sensitive to
such low energies. 

We have determined the expected fluorescence flux observed on ground taking into account also the extinction due to the atmosphere. A comparison of the calculated flux with the background at the detection site provides an immediate limit to the detection capabilities of a GRB signal. This limit, or threshold, can be expressed as a fraction 
$\delta$ of the background photon flux. 

In order to estimate the expected GRB detection rate we have used the
Swift data \cite{SWIFT} extrapolated to higher photon intensities. It
has been shown that, only in the case of a fluorescence detection
sensitive to an excess above one percent of the background flux, it is
possible to detect at least one GRB in ten years, fixing the most
favorable conditions in the Swift data extrapolation. The ability of
reaching such or better sensitivities has been discussed on general grounds. 

Limits on fluence achievable by surface experiments are of the order of 10$^{-5}$ erg/cm$^2$ or better.
We can reasonably expect that whatever threshold could be achieved in fluorescence detection,
its overall sensitivity is worse by some order of magnitude with respect to other on-ground techniques,
unless dedicated detectors and/or more selective algorithms are used. Conversely, as already pointed out,
the fluorescence detection does not suffer from the same limitation of the other ground techniques, i.e. the actual continuation of the GRB spectrum up to the 10 GeV region, which is not yet demonstrated. 

\section*{Acknowledgments}
We are grateful to Xavier Bertou and Paul Sommers for very helpful discussions. The work of RA and AFG is partially funded by the contract 
ASI-INAF I/088/06/0 for theoretical studies in High Energy Astrophysics. 
The work of CM and FS is partially funded by  
ALFA-EC/HELEN.

\end{document}